\documentclass[sigconf]{cidr-2027}

\usepackage{caption}
\usepackage{pifont}
\definecolor{lin}{HTML}{e31a1c}

\definecolor{donna}{HTML}{1f78b4}

\newcommand{\indexName}{Periplus}

\begin{document}
\title{Rethinking Document Indexing for LLM-based Analysis}
\author{Donna Pham, Zhongwei Xu, Siyuan Dong, Haotian Gong, Lin Ma}
\affiliation{%
  \institution{University of Michigan}
  \city{Ann Arbor}
  \state{Michigan}
  \country{USA}
}
\email{{phdonn,xzw,dougdong,haotiang,linmacse}@umich.edu}

\begin{abstract}
Documents such as clinical records, legal contracts, and scientific papers are increasingly analyzed with large language models (LLMs). The naïve approach of prompting the model with the full text for every question is slow, expensive, prone to hallucination on long inputs, and reuses no work across questions. Retrieval-augmented generation (RAG) systems build reusable state to address this, but they store it as embedded chunks: slicing text into fragments can lose the meaning that spans them, and the resulting index is tied to one embedding model, so it must be rebuilt whenever that model changes. Answering many questions over a large body of data without a full rescan, however, is not a new problem: it is the problem database indexes have been solving for decades, and we argue that this accumulated index design knowledge is waiting to be revisited for LLM access.

As a concrete starting point, we propose Periplus, a hierarchical plain-text index inspired by the classic B+ tree. It arranges a document into pages running from general summaries at the root to specific ones at the leaves, so that each level guides the search into the next. But it departs from the B+ tree in three ways suited to LLM access: content lives at every level, the pages are plain text rather than attribute values, and navigation follows the relevance of a page summary to the question rather than a key comparison. The structure itself is discovered per document by the LLM rather than hand-designed in advance. In a preliminary evaluation on narrative and contract corpora, Periplus comes within 0.1--0.9 percentage points of the strongest baseline's accuracy while consuming 70--72\% fewer total tokens. These early results are encouraging, and they offer only a glimpse of what index design can bring to LLM-based analysis.

\end{abstract}

\maketitle

\begingroup
\renewcommand{\thefootnote}{}
\footnotetext{\textbf{Code Availability.} The \indexName{} implementation is publicly available at \url{https://github.com/umich-db/Periplus}.}
\endgroup

\section{Introduction}
\label{sec:intro}

\begin{figure}[t]
  \centering
  \includegraphics[width=0.88\columnwidth]{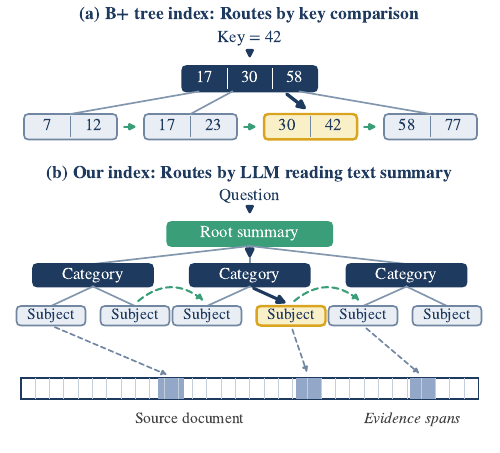}
  \caption{(a) A B+ tree routes a lookup by comparing a search key against separator keys, and content lives only at the leaves. (b) Periplus preserves the hierarchy but replaces separator keys with plain-text summaries and key comparison with an LLM reading them. Content lives at every level, and each subject points back to its evidence spans in the source document.}
  \label{fig:headline}
\end{figure}

Documents such as clinical records, legal contracts, and scientific papers are routinely examined in depth, and analyzing them by hand is slow and labor-intensive. A recent and promising line of work uses large language models (LLMs) to do this automatically, answering the free-form questions that previous keyword and extraction tools could not~\cite{patel2025lotus,shankar2025docetl}.

However, prompting an LLM with the full document on every question incurs a token cost linear in document length, compounding when the same document is analyzed repeatedly: a patient record consulted over time, a contract revisited across legal matters, a scientific paper queried for distinct analyses. Long contexts also degrade accuracy, as models miss evidence buried in long inputs even when it is present~\cite{liu2024lost}. Recent systems decompose execution into agentic pipelines (DocETL~\cite{shankar2025docetl}) or declarative queries (Lotus~\cite{patel2025lotus}, Palimpzest~\cite{liu2025palimpzest}) over text chunks, yet they remain \emph{stateless}: each query is processed from scratch over the raw source text, paying the full cost of ingestion again on every query.

A natural optimization is to precompute an intermediate state once for reuse across queries, which we broadly categorize into two families of approaches.


\textbf{The first family builds a vector index over the document:} RAG~\cite{lewis2020rag} embeds chunks as vectors, and GraphRAG~\cite{edge2024graphrag} and LightRAG~\cite{guo2025lightrag} add graph structure for whole-document questions. But chunking severs connections between distant passages, top-$k$ retrieval surfaces only a few, and the index is tied to one embedding model, forcing a rebuild whenever it changes. \textbf{The second family supports structured document queries:} Evaporate~\cite{arora2023evaporate} and Doctopus~\cite{chai2025doctopus} extract targeted attributes into tables. ZenDB~\cite{lin2024zendb} takes a hierarchical approach, building a tree from document sections and headers with summaries at each node to guide query-time retrieval for SQL analytics. ZenDB also supports lazily specified attributes rather than requiring all attributes to be extracted upfront. However, its tree organization follows the document's existing section structure, and its query interface operates over user-defined relational tables. This makes it less direct to ask open-ended questions that require connecting information across sections without first expressing the analysis through a structured schema.

In this paper, we explore a different path. \textit{We return to the successful principles from the database literature, where answering many queries over a large body of data without rescanning it is precisely the problem that the index, and in particular the B+ tree~\cite{comer1979ubiquitous, bayer1972organization}, has solved for decades.} We present \textbf{\indexName{}}, a prototype \textbf{hierarchical, plain-text index inspired by the B+ tree} that any LLM can read directly (Figure~\ref{fig:headline}): a high-level summary at the root, category-level summaries beneath it, and subjects with detailed fields at the leaves. \indexName{} is built once for the target document(s) and consulted by every subsequent analysis, just as a database index is built once and reused for every query. Similar to a B+ tree, which sizes its nodes to a disk page, \indexName{} sizes them to the answering model's context window.

\indexName{} borrows the B+ tree's shape and hierarchical access over pages, but departs from it in three ways suited to LLM access. Content lives at every level, not only at the leaves, so many questions are answered from an internal summary without descending further, much as a covering index answers a query without touching the base table~\cite{graefe2011btree}. The pages are plain text rather than embedded vectors, so any model can read them, context across distant passages is preserved rather than isolated in chunks, and nothing needs rebuilding when models change. And navigation is semantic: the model judges which summaries are relevant rather than comparing a search key, and because the organization is discovered from document content rather than defined by its existing section hierarchy or a user-specified relational schema, \indexName{} supports deep, open-ended analysis without requiring predefined query attributes.

Realizing this vision raises a range of research questions.
As an initial exploration, we design and implement \indexName{}, a prototype index for a single document, covering how its structure is discovered, how the document is compiled into it, how the resulting tree is balanced, and how queries are answered against it (Section~\ref{sec:method}).
In a preliminary evaluation on movie screenplays (NarrativeQA~\cite{kocisky2018narrativeqa}) and legal contracts (CUAD~\cite{hendrycks2021cuad}), \indexName{} comes within 0.1--0.9 percentage points in accuracy of the strongest baseline, DocETL, while consuming 70--72\% fewer total tokens (Section~\ref{sec:eval}).

Much of the machinery that makes database indexes practical remains to be worked out here: updates, concurrent access, cheaper construction and lookup, and extension from one document to a corpus. Nevertheless, our preliminary results suggest classical index design is a productive lens for LLM-based document analysis: \indexName{} already approaches the accuracy of the strongest baseline at a fraction of its token cost, and this paper is an early step toward revisiting decades of index design knowledge for this setting.

\section{Related Work}
\label{sec:background}

We categorize existing approaches to LLM-based document analysis into three distinct paradigms based on how they manage state, execution cost, and schema flexibility.

\textbf{Stateless Execution Pipelines.}
The most straightforward approach evaluates each query directly over the raw text. Advanced declarative systems decompose complex queries into sequences of LLM operators over document chunks: DocETL~\cite{shankar2025docetl} does so with agentic query rewriting, extended by interactive steering~\cite{shankar2025docwrangler} and multi-objective optimization~\cite{wei2026multiobjective}, while Lotus~\cite{patel2025lotus} and Palimpzest~\cite{liu2025palimpzest} optimize similar declarative plans. Flexible and schema-free, these pipelines are \emph{stateless} but every query re-pays the full reading cost, and long inputs raise hallucination rates~\cite{liu2024lost,roig2026hallucinate}.

\textbf{Vector, Graph, and Tree Indexes.}
A second paradigm precomputes reusable state over text chunks. RAG~\cite{lewis2020rag} retrieves the top-$k$ embedded chunks, GraphRAG~\cite{edge2024graphrag} and LightRAG~\cite{guo2025lightrag} layer entity graphs and community summaries over them for global queries, and LEANN~\cite{wang2025leann} prunes the vector index to cut storage, targeting footprint rather than retrieval quality. Among these methods, RAPTOR~\cite{sarthi2024raptor}, which recursively clusters and summarizes chunks into a tree of summaries, has the closest layout to \indexName{}. RAPTOR fixes one set of choices in a much larger space: how the hierarchy is constructed, what each node stores, and how a query descends it are all open design decisions. The database literature has explored precisely these decisions for key-based access, and we argue that this knowledge should inform index design for LLM-based analysis; \indexName{}, described in Section~\ref{sec:method}, is one point in that space.


\textbf{Structured Information Extraction.}
A third paradigm extracts targeted attributes into relational tables before query time, as in Evaporate~\cite{arora2023evaporate} and Doctopus~\cite{chai2025doctopus}, which selects budget-optimal extraction strategies per attribute. These approaches work well for known attributes, but their fixed extraction schemas limit questions requiring unanticipated information or synthesis across passages.
ZenDB~\cite{lin2024zendb} instead builds a Semantic Hierarchical Tree (SHT) from document sections and headers, using node summaries to locate relevant passages and supporting lazy attribute extraction for SQL queries. Unlike ZenDB's section-based organization and relational query interface, Periplus uses LLM-discovered semantic categories and evidence-linked text pages to support free-form question answering.

\section{Method}
\label{sec:method}


\begin{figure*}[t]
  \centering
  \includegraphics[width=0.86\textwidth]{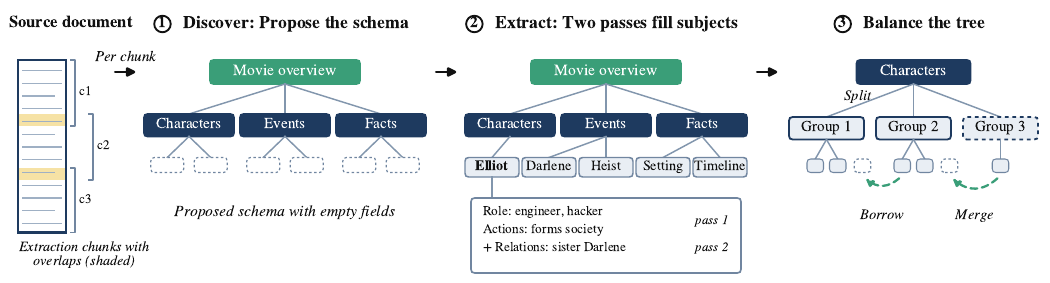}
  \caption{Compiling a document into \indexName{}. The document is split into overlapping chunks; (1) discovery proposes a schema and a partial overview per chunk and merges them; (2) two extraction passes over the chunks fill the schema, each extracted field recording its evidence span; (3) the balancing pass splits, borrows, and merges subject groups against occupancy bounds.}
  \label{fig:compile}
\end{figure*}

\subsection{Overview}
\label{sec:overview}
We take the B+ tree as our starting point and ask which of its elements remain applicable when the index must be read by a model rather than searched by key. \indexName{} is a hierarchical, plain-text index compiled once per document and consulted by every later query (Figure~\ref{fig:headline}b). We describe it following the conventional presentation of a B+ tree: first its structure, then the invariants it maintains, and finally the operations that manage them.

\paragraph{Structure}
The tree has three types of nodes (root, internal, and leaf). At the top, a \emph{root} holds a dense, consolidated summary of the entire document and routes queries to the internal level. Internal nodes come in two forms: \emph{categories}, the broad divisions of content proposed at discovery, and \emph{grouping} nodes, inserted beneath a category when it holds more subjects than a single node may hold, so depth grows only where content forces it. At the leaf level sit \emph{subjects}: the concrete entities or thematic units of the text, such as a character in a screenplay, a clause in a contract, or an experimental procedure in a paper. Each subject records a small set of \emph{fields} extracted directly from the source, and every field is linked to the evidence span that produced it. The tree is therefore at least three levels deep, and deeper wherever grouping nodes are inserted. Every node, at every level, is stored as human-readable plain text.

\paragraph{Invariants}
A classical B+ tree of order $b$ keeps every internal node between $\lceil b/2 \rceil$ and $b$ children, so search visits $O(\log_b n)$ nodes. \indexName{} keeps invariants of a similar shape, parameterized instead by two budgets on what a query reads: a \emph{routing budget}, the most one routing step may read while choosing a direction (Section~\ref{sec:balance}), and an \emph{answering budget}, the total evidence a query may collect (Section~\ref{sec:query}). A node's summary may not exceed a capacity limit (overflowing nodes split), and its child count stays inside a fanout window enforced by split, borrow, and merge (Section~\ref{sec:balance}); with fanout bounded on both sides, a descent still reads only logarithmically many routing pages. We depart from the classical invariants in two places: content lives at every level, so a query may be answered before reaching a leaf, and balance is local to each category rather than global, since forcing global symmetry onto unevenly distributed content degraded accuracy (Section~\ref{sec:discovery}). LLM inference is performed at build time (discovery, extraction, revision) and query time (routing the tree traversal and synthesizing the answer).

\paragraph{Operations}
We design four operations for \indexName{}. \emph{Discovery} proposes the document's schema (Section~\ref{sec:discovery}). \emph{Compilation} fills it (Section~\ref{sec:compile}). \emph{Balancing} restores the occupancy invariants (Section~\ref{sec:balance}). \emph{Query answering} descends the tree (Section~\ref{sec:query}). Other index operations, including insert, delete, and concurrent access, are not yet built; we discuss them in Section~\ref{sec:extensions}.


\subsection{Discovering the Structure}
\label{sec:discovery}
Discovery proposes the document's schema before any content is extracted (Figure~\ref{fig:compile}, Step (1)).

\paragraph{Cleaning and Chunking}
The raw text is first stripped of download artifacts and markup, then split into sequential chunks of about 30{,}000 characters that overlap by 2{,}000.
Chunks are large enough to keep the number of LLM calls small while still fitting comfortably within the context window, and the overlap ensures that a passage cut at one chunk's edge appears in full in the adjacent one.
Compilation later reuses this same chunking (Section~\ref{sec:compile}).

\paragraph{Proposing the Schema}
Discovery runs one LLM call per chunk, each proposing the categories it sees, the fields worth recording under each, and a partial overview.
The proposals are merged into a single schema: categories with short descriptions and field templates, plus an initial overview that seeds the root. On a screenplay, this might yield \textit{characters}, \textit{events}, and \textit{facts}, with fields under \textit{characters} such as \textit{role}, \textit{actions}, and \textit{relationships}; nothing is templated in advance.
Merging uses fixed rules: repeated entries are deduplicated by content, and the first non-empty description is kept.
This leaves an open question of how to reconcile entries that are near-duplicates rather than exact ones.
Discovery thus fixes only the structural types and a provisional root overview; the concrete subjects, field values, and evidence spans arrive during compilation, which also revises the overview once extraction completes (Section~\ref{sec:compile}).
In our current implementation, each subject page is capped at 2{,}400 characters; content beyond the cap moves under a grouping node with its own summary, as writing describes (Section~\ref{sec:compile}). Sizing this cap dynamically against extracted volume is an open direction.

\paragraph{Top-Down Schema vs. Bottom-Up Clustering}
\indexName{} fixes the categories top-down, before any leaf is grouped.
The alternative is to build the index bottom-up, similar to the classic method to bulk-load a B+ tree from sorted keys~\cite{graefe2011btree}.
We tested a variant that forms levels by recursively clustering embedding similarities across leaf passages: it grouped similar subjects more cleanly, yet accuracy fell from 55.5\% to 47.1\%.
Keys carry a global order, and text carries none, so deciding what belongs together requires a holistic reading of the document.
Local similarity clusters nearby facts but scatters the context that jointly answers a single question.
Whether a bottom-up build can work remains an open question; within a category, where the top-level decision is already fixed, we still use embedding similarity to group subjects during balancing (Section~\ref{sec:balance}).

\subsection{Compiling a Document}
\label{sec:compile}
After discovery, compilation reads every chunk and fills in the schema in three steps: extract, revise, and write (Figure~\ref{fig:compile}, Step (2)).

\paragraph{Extracting with Two Passes}
Extraction runs two passes over the chunks.
The first pass reads each chunk against the discovered schema and adds the subjects it finds (e.g., specific characters or events) to a running merged record; categories remain fixed while subjects accumulate.
Reading each chunk in isolation, however, the first pass cannot connect a passage to a subject that a different chunk introduced.
The second pass therefore revisits each chunk with the merged record in context, attaching such passages to the subjects they describe, filling still-empty fields, and recovering details that looked incidental on an isolated read.
Carrying the record in context requires it to fit within the model's context window, which holds on the corpora we evaluate.

The model returns each field as either a list or a single value, and merging handles the two cases differently.
List fields, such as a character's \textit{actions}, accumulate entries with exact duplicates removed.
Single-value fields, such as a character's \textit{role}, keep the first non-empty value, on the assumption that a document defines a fact where it first states it; resolving conflicting values more comprehensively is left to future work.

Compressing a document into a schema inevitably leaves things out. When a subject was extracted but its page omits a detail a later question needs, that detail is still reachable: every extracted field carries its exact evidence span, so the query can fall back to the originating chunk (Section~\ref{sec:query}). Content that fits no category at all is not recovered this way, and remains a limitation of compilation.

\paragraph{Revising the Root}
After the second pass, one further LLM call rewrites the root's provisional overview against the full merged record, since the provisional draft saw only per-chunk glimpses; the result is a dense, document-wide summary every query begins reading first (Section~\ref{sec:query}).

\paragraph{Writing}
Writing materializes \indexName{} into plain-text files: one overview file at the root summarizing the schema, and one file per subject under its parent category, matching the hierarchy in Figure~\ref{fig:headline}b. Each page is capped at 2{,}400 characters. When a category holds more subjects than one page can list, writing inserts \emph{grouping nodes} to hold the overflow, keeping children in the order extraction produced them. Writing also records the short \emph{routing descriptor} for the root and each internal node for tree traversal, and the links between related subjects that queries follow laterally (Section~\ref{sec:query}).

\subsection{Balancing the Tree}
\label{sec:balance}
The tree at this point is at least three levels deep (root, categories, subjects), plus the grouping nodes inserted at write time. It can also be imbalanced because content extracted from the document may not be evenly distributed: a screenplay may yield dozens of characters, but only a few settings, and the widest node in our corpus has 73 children. To keep every routing step within the routing budget, a balancing step reshapes the raw hierarchy from the compilation step, normalizing the depth and fanout that earlier phases produced unevenly (Figure~\ref{fig:compile}, Step (3)).

\paragraph{Deriving Occupancy Bounds}
Both occupancy bounds derive from the routing budget $C$. If a parent routing node allocates at most $C$ characters and each child routing descriptor consumes $S$ characters, the maximum node fanout $f$ is capped at $f = \lfloor C/S \rfloor$. In our configuration ($C = 4{,}000$ characters, $S = 120$ characters), fanout is capped at $f = 33$, and, following textbook B+ tree occupancy rules, non-root internal nodes are kept within the window $[\lceil f/2 \rceil, f] = [17, 33]$. The descriptor length $S$ is itself a design knob with a trade-off: longer descriptors give the router more to read but reduce the fanout, thereby deepening the tree.

\paragraph{Splitting via 1D Semantic Projection}
A split triggers whenever a node holds more than $f$ children. The write-time groupings bound page size, not fanout, and take children in arrival order, so wide nodes survive compilation; splitting is what bounds fanout, and it chooses group membership semantically. Classical B+ trees sort scalar keys along a 1D axis, but text summaries have no natural scalar order. To establish one without LLM calls, each child's routing descriptor is embedded with a local model and projected onto the children's first principal component, so similar children sit next to each other. The ordered list is cut into contiguous, near-equal groups, each becoming a new grouping node, and the split recurses inside any group that still exceeds $f$, so a very wide category may gain more than one grouping level.

\paragraph{Restoring Occupancy via Borrow and Merge}
Splitting bounds fanout from above, but some nodes sit below the floor $\lceil f/2 \rceil$. These are mostly the remainder groups that write-time grouping leaves behind.
An underfull group first borrows, adopting the most similar children from a sibling that can give them up without itself dropping below the floor; if no sibling can spare any, the group merges with its most similar sibling.
The pass traverses from the deepest level upward, since repairing one level changes how many children its parent holds and can push the parent out of range in turn, and repeats until no node violates the bounds.
A category whose total subject count is below the floor is left alone. This preserves the domain hierarchy discovered by the LLM: forcing global symmetry across unevenly distributed categories degrades answer accuracy (Section~\ref{sec:discovery}).
The pass uses only local embeddings, which are discarded once the build completes: nothing but plain text is stored, so \indexName{} needs no rebuild when the reading model changes.

\subsection{Answering Queries}
\label{sec:query}
A query is answered by orienting at the root and descending the tree, falling back to the source text when \indexName{} cannot supply the answer.
The model synthesizes answers only from the collected evidence, abstaining when the required facts are missing.

\paragraph{Orientation \& Semantic Traversal}
Execution pairs the query with the root summary first, so the model has a document-wide overview, then descends through category, grouping, and subject nodes.
At each level, a short LLM call reads only the children's routing descriptors and judges which are relevant, so a descent costs a handful of brief calls rather than one long read of the document.
Descent is not the only way to move.
Just as a B+ tree links its leaves so that a search can continue sideways instead of returning to the root, subjects carry links to related subjects, which may sit under other categories (e.g., a character's page links to the events that character appears in).
A query that lands on a subject can therefore follow these links laterally rather than descend again from the root when the model judges a linked subject relevant to the question. Selected pages accumulate in a buffer capped at the answering budget (12{,}000 characters), which bounds how far a query can roam, and is then handed to the model with the instruction to return a concise answer or output exactly \emph{not found}. Setting this budget adaptively, rather than fixing it in advance, is an open direction.

\paragraph{Source-Grounded Fallback}
A \emph{not found} from Periplus triggers the fallback: sometimes it points to the right subject, but its page omits the specific detail the question asks for. Using the evidence spans recorded during compilation (Section~\ref{sec:compile}), the fallback returns to the originating chunk and answers from the source text directly (instead of rescanning the full document). On the corpora we evaluate, the fallback fired on 18\% of questions. When the source chunk also does not contain the answer, the query returns \emph{not found}.

\subsection{Architectural Extensions and Future Work}
\label{sec:extensions}
Beyond the refinements noted above, a few index operations remain unbuilt.
The first is maintaining Periplus as its document changes: amended contracts and revised papers call for an \emph{insert} that folds new content into the tree, and for a \emph{deletion} that removes what has been superseded.
Neither is as mechanical as its classical counterpart, since a plain-text index must first decide which subjects the new text belongs to, and whether it warrants reopening the schema that discovery fixed.
A second question is concurrency: when many readers and writers share one index, as in interactive legal or clinical analysis, what does latch crabbing look like for pages whose summaries a writer may rewrite mid-traversal?
A third challenge is lookup latency: traversing Periplus requires visiting index pages sequentially, with an LLM call at each level to determine the next page, potentially adding substantial end-to-end delay.
Reducing these sequential calls through caching, parallel traversal, or more efficient routing is an important direction for future work.
Finally, Periplus should extend from a single document to a corpus. Lifting each document's root summary into a higher-level node is the obvious construction, but whether the right organization follows document boundaries, rather than merging related documents or splitting long ones, is an open question.

\begin{figure}[t]
  \centering
  \includegraphics[width=\columnwidth]{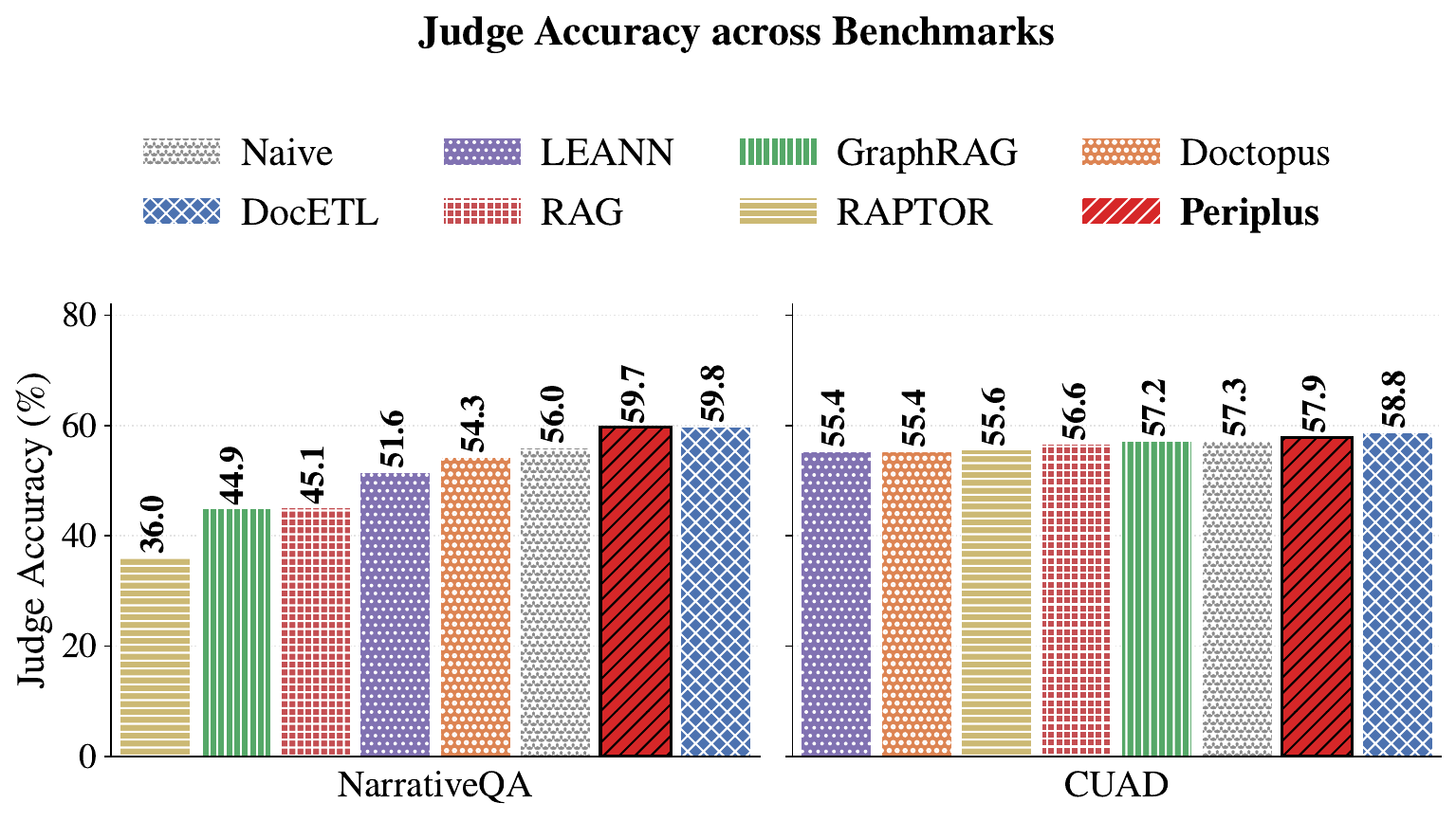}\\
  {\footnotesize (a) Judge accuracy}\\[4pt]
  \includegraphics[width=\columnwidth]{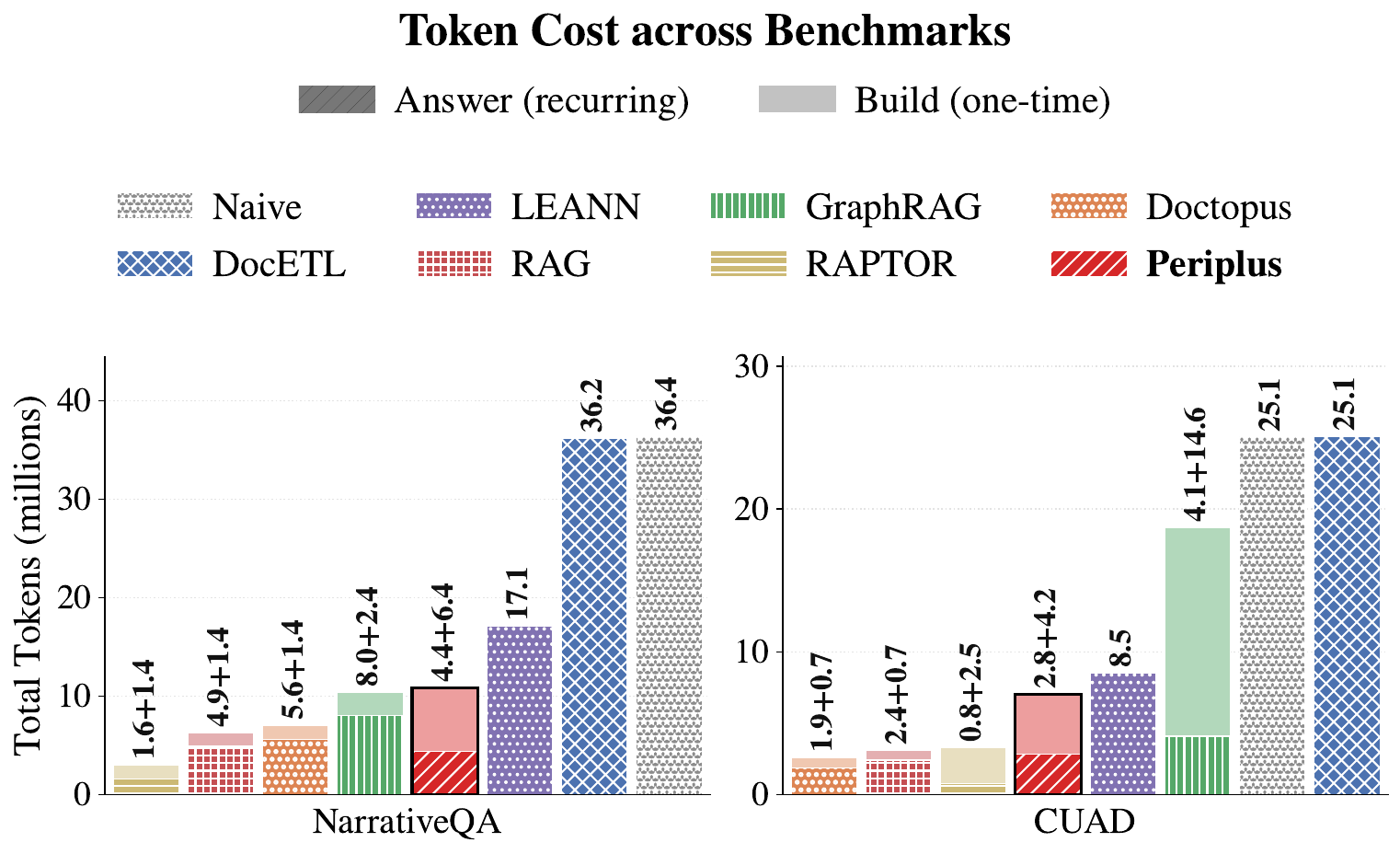}\\
  {\footnotesize (b) Input tokens, one-time build plus recurring answering}
  \caption{Measured results on NarrativeQA (885 free-form questions over 30 movie screenplays) and CUAD (410 clause questions over 10 legal contracts).}
  \label{fig:results}
\end{figure}

\section{Preliminary Evaluation}
\label{sec:eval}

\subsection{Experimental Setup}

\begin{figure}[t]
  \centering
  \begin{minipage}[t]{0.9\columnwidth}
    \centering
    \includegraphics[width=\linewidth]{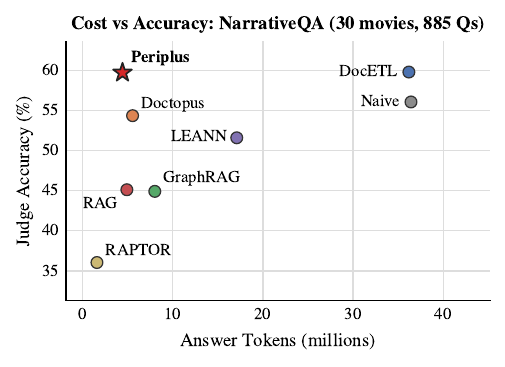}\\[-1pt]
    {\footnotesize (a) NarrativeQA}
  \end{minipage}\hfill
  \begin{minipage}[t]{0.9\columnwidth}
    \centering
    \includegraphics[width=\linewidth]{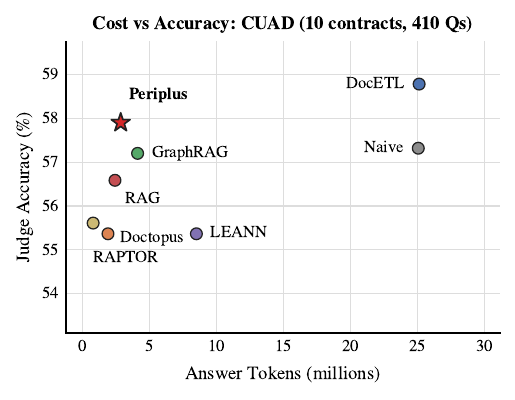}\\[-1pt]
    {\footnotesize (b) CUAD}
  \end{minipage}
  \caption{Answer cost versus judge accuracy. (a) NarrativeQA. (b) CUAD. Each point is an approach; the star marks Periplus. On both corpora, it sits near the top of the accuracy range at a fraction of the answering cost of the stateless systems.}

  \label{fig:tradeoff}
\end{figure}

We conduct a preliminary evaluation on two datasets with different structures: thirty movie screenplays from \textbf{NarrativeQA}~\cite{kocisky2018narrativeqa} (885 free-form questions; hundreds of pages each, answers requiring synthesis across distant scenes) and ten annotated legal contracts from \textbf{CUAD}~\cite{hendrycks2021cuad} (410 clause questions; more regular structure).

GPT-5~\cite{openai2025gpt5} serves as the index-building and answering model, while GPT-6.1 Sol acts as the judge that scores answer equivalence against the references. We compare against representatives of each family from Section~\ref{sec:background}: a full-document baseline (\emph{Naive}: the entire document fed to the model per question), the stateless DocETL pipeline~\cite{shankar2025docetl}\footnote{MOAR~\cite{wei2026multiobjective} optimizes a DocETL pipeline for repeated execution; our workload targets a different scenario where each question is asked once, so we run DocETL without it.}, the structured-extraction system Doctopus~\cite{chai2025doctopus}, and four vector-index baselines: RAG~\cite{lewis2020rag}, GraphRAG~\cite{edge2024graphrag}, RAPTOR~\cite{sarthi2024raptor}, and LEANN~\cite{wang2025leann}.

We report \textbf{accuracy}, the fraction of answers the judge deems equivalent to the reference, and \textbf{cost}, total input tokens split into one-time build and recurring answering. On CUAD's many no-answer questions, a system is correct only if it responds that the contract does not contain the answer; we check this by looking for phrases like \emph{not found} in its output, the same check for every system, instead of asking the judge. All numbers come from complete, automated runs with no manual intervention. To validate the judge, we blind-labeled a stratified sample of 112 verdicts (seven per method per dataset) and agreed with it in 98\% of cases.

\subsection{Results}

\noindent\textbf{Accuracy on NarrativeQA.} Figure~\ref{fig:results}a gives the ranking on NarrativeQA. Periplus reaches 59.7\% judge accuracy, second only to DocETL (59.8\%), a gap of just 0.1 percentage points. DocETL re-processes the full document with agentic rewriting on every question; instead, Periplus incurs a one-time pre-processing cost at build time. The Naive baseline ranks third at 56.0\%, followed by Doctopus at 54.3\%; NarrativeQA's questions concentrate on characters and events that a well-chosen extraction schema can capture. The vector-index baselines all fall below Naive: a top-$k$ slice of a hundreds-of-pages screenplay often omits the scenes a free-form question depends on, while Periplus retains those cross-scene connections in its structure. The inaccurate answers from Periplus concentrate on \emph{why} and \emph{how} questions, where compression drops the connective detail those questions depend on, even as the pages preserve what happened; sizing pages against the query workload (Section~\ref{sec:compile}) is a potential optimization to keep that detail where questions demand it.

\medskip
\noindent\textbf{Accuracy on CUAD.}
On legal contracts, the ranking changes. Periplus reaches 57.9\%, again second to DocETL (58.8\%) and within 0.9 percentage points. The vector-index systems perform better on contracts: GraphRAG rises from 44.9\% to 57.2\%, LEANN from 51.6\% to 55.4\%, and RAG from 45.1\% to 56.6\%. CUAD's questions are local and templated, which chunked retrieval handles well. Doctopus scores 55.4\% on CUAD. Its edge is organizing scattered facts into a table, and a contract arrives already organized into labeled clauses. Here, Periplus's misses focus on exact-value questions, where consolidation keeps a clause's heading but drops the term within it; better subject-level traversal and better-calibrated fallback to source may help. Across both corpora, retrieval-based indexes are stronger on the structured contracts, while extraction is more competitive on narratives. Periplus chooses the organization per document and is the only indexed system in the top two on both datasets.

\medskip
\noindent\textbf{Token cost.}
Figure~\ref{fig:results}b separates one-time build cost from recurring answering cost. Stateless systems pay everything at answer time: Naive spends 36.4M tokens on NarrativeQA and DocETL 36.2M, roughly 41K per question, and both spend ${\sim}25$M again on CUAD. Periplus inverts the pattern: 6.4M to build and 4.4M to answer all 885 NarrativeQA questions (about 5K each), 4.2M and 2.8M on 410 CUAD questions (about 7K each). This amounts to 70\% fewer total input tokens than DocETL on NarrativeQA and 72\% fewer on CUAD. The remaining index systems achieve cheaper builds (0.7--2.5M) at lower accuracy, with two exceptions: GraphRAG's graph construction costs 14.6M on CUAD, 3.5$\times$ Periplus's build, for accuracy just below Periplus, and LEANN defers embedding to query time, making its build nearly free but its answering the most expensive of any index system (17.1M on NarrativeQA). Periplus's one-time build cost is repaid quickly: at 5K per question against DocETL's 41K, the NarrativeQA build recovers after roughly 180 questions, six per screenplay, and the CUAD build after roughly 80, eight per contract, both within the first third of the measured workload; every question thereafter costs about an eighth of the stateless alternative (Figure~\ref{fig:tradeoff}).

\section{Conclusion}
\label{sec:conclusion}

This paper has argued that the design knowledge accumulated around database indexes can be applied to LLM-based document analysis, and has presented one such design, \indexName{}: a hierarchical plain-text index in which an LLM discovers the schema, extraction fills it, and a balancing pass enforces occupancy bounds derived from the routing budget.

It comes within 0.1 percentage points of DocETL on NarrativeQA and 0.9 percentage points on CUAD while answering at roughly an eighth of the tokens, with a build cost repaid after six questions per screenplay and eight per contract.

Much of the machinery that makes database indexes practical remains to be revisited: how to update the index when the document changes, how to coordinate concurrent readers and writers over a shared index, and how to extend the index from a single document to a corpus.

We offer this work as an early entry in that design space.

\begin{acks}
This work was supported (in part) by the Google ML and Systems Junior Faculty Award, the Google JAX AI Stack Research Award, the NVIDIA Academic Grant Program Award, the AWS Agentic AI Amazon Research Awards, and the University of Michigan’s Strategic Initiative Fund.
\end{acks}

\bibliographystyle{ACM-Reference-Format}
\bibliography{sample}

\end{document}